\documentclass[aps,prl,twocolumn,superscriptaddress,showpacs,preprintnumbers,amssymb,tightenlines,floatfix]{revtex4}
\usepackage[fleqn]{amsmath}
\usepackage{graphicx} 
\usepackage{dcolumn}  
\usepackage{xcolor}
\usepackage{subfigure}
\usepackage{mathtools}
\usepackage{units}
\graphicspath{{ps}}

\begin{document}

\title{ \quad\\[0.5cm]  Measurement of the inclusive semileptonic branching
fraction  $\mathcal{B}(B_s^0 \to X^- \ell^+ \nu_\ell)$ at Belle}

\begin{abstract}
We report a measurement of the inclusive semileptonic $B_s^0$ branching fraction
in a $121$~fb${}^{-1}$ data sample collected near the $\Upsilon(5S)$ resonance 
with the Belle detector at the KEKB asymmetric energy $e^+e^-$ collider. Events 
containing  $B_s^{0(*)} \bar{B}_s^{0(*)}$ pairs are selected by reconstructing a tag side 
$D_s^+$ and identifying a signal side lepton $\ell^+$ ($\ell = e, \mu$) that is
required to have the same-sign charge to ensure that both originate from different
$B_s^0$ mesons. The $B_s^0 \to X^- \ell^+ \nu_\ell$
branching fraction is extracted from the ratio of the measured yields of $D_s^+$ mesons and
$D_s^+\ell^+$ pairs and the known production and branching fractions.
The inclusive semileptonic branching fraction is measured to be
$[9.6 \pm 0.4(\text{stat}) \pm 0.7(\text{syst})]\%$.
\end{abstract}

\affiliation{University of the Basque Country UPV/EHU, 48080 Bilbao}
\affiliation{University of Bonn, 53115 Bonn}
\affiliation{Budker Institute of Nuclear Physics SB RAS and Novosibirsk State University, Novosibirsk 630090}
\affiliation{Faculty of Mathematics and Physics, Charles University, 121 16 Prague}
\affiliation{University of Cincinnati, Cincinnati, Ohio 45221}
\affiliation{Gyeongsang National University, Chinju 660-701}
\affiliation{Hanyang University, Seoul 133-791}
\affiliation{University of Hawaii, Honolulu, Hawaii 96822}
\affiliation{High Energy Accelerator Research Organization (KEK), Tsukuba 305-0801}
\affiliation{Hiroshima Institute of Technology, Hiroshima 731-5193}
\affiliation{Ikerbasque, 48011 Bilbao}
\affiliation{Indian Institute of Technology Guwahati, Assam 781039}
\affiliation{Indian Institute of Technology Madras, Chennai 600036}
\affiliation{Institute of High Energy Physics, Chinese Academy of Sciences, Beijing 100049}
\affiliation{Institute of High Energy Physics, Vienna 1050}
\affiliation{Institute for High Energy Physics, Protvino 142281}
\affiliation{Institute for Theoretical and Experimental Physics, Moscow 117218}
\affiliation{J. Stefan Institute, 1000 Ljubljana}
\affiliation{Kanagawa University, Yokohama 221-8686}
\affiliation{Institut f\"ur Experimentelle Kernphysik, Karlsruher Institut f\"ur Technologie, 76131 Karlsruhe}
\affiliation{Korea Institute of Science and Technology Information, Daejeon 305-806}
\affiliation{Korea University, Seoul 136-713}
\affiliation{Kyungpook National University, Daegu 702-701}
\affiliation{\'Ecole Polytechnique F\'ed\'erale de Lausanne (EPFL), Lausanne 1015}
\affiliation{Faculty of Mathematics and Physics, University of Ljubljana, 1000 Ljubljana}
\affiliation{Luther College, Decorah, Iowa 52101}
\affiliation{University of Maribor, 2000 Maribor}
\affiliation{Max-Planck-Institut f\"ur Physik, 80805 M\"unchen}
\affiliation{School of Physics, University of Melbourne, Victoria 3010}
\affiliation{Moscow Physical Engineering Institute, Moscow 115409}
\affiliation{Moscow Institute of Physics and Technology, Moscow Region 141700}
\affiliation{Graduate School of Science, Nagoya University, Nagoya 464-8602}
\affiliation{Kobayashi-Maskawa Institute, Nagoya University, Nagoya 464-8602}
\affiliation{Nara Women's University, Nara 630-8506}
\affiliation{National Central University, Chung-li 32054}
\affiliation{National United University, Miao Li 36003}
\affiliation{Department of Physics, National Taiwan University, Taipei 10617}
\affiliation{H. Niewodniczanski Institute of Nuclear Physics, Krakow 31-342}
\affiliation{Nippon Dental University, Niigata 951-8580}
\affiliation{Niigata University, Niigata 950-2181}
\affiliation{Osaka City University, Osaka 558-8585}
\affiliation{Pacific Northwest National Laboratory, Richland, Washington 99352}
\affiliation{Research Center for Electron Photon Science, Tohoku University, Sendai 980-8578}
\affiliation{University of Science and Technology of China, Hefei 230026}
\affiliation{Seoul National University, Seoul 151-742}
\affiliation{Sungkyunkwan University, Suwon 440-746}
\affiliation{School of Physics, University of Sydney, NSW 2006}
\affiliation{Tata Institute of Fundamental Research, Mumbai 400005}
\affiliation{Excellence Cluster Universe, Technische Universit\"at M\"unchen, 85748 Garching}
\affiliation{Toho University, Funabashi 274-8510}
\affiliation{Tohoku Gakuin University, Tagajo 985-8537}
\affiliation{Tohoku University, Sendai 980-8578}
\affiliation{Department of Physics, University of Tokyo, Tokyo 113-0033}
\affiliation{Tokyo Institute of Technology, Tokyo 152-8550}
\affiliation{Tokyo Metropolitan University, Tokyo 192-0397}
\affiliation{Tokyo University of Agriculture and Technology, Tokyo 184-8588}
\affiliation{CNP, Virginia Polytechnic Institute and State University, Blacksburg, Virginia 24061}
\affiliation{Wayne State University, Detroit, Michigan 48202}
\affiliation{Yamagata University, Yamagata 990-8560}
\affiliation{Yonsei University, Seoul 120-749}

 \author{C.~Oswald}\affiliation{University of Bonn, 53115 Bonn} 
 \author{P.~Urquijo}\affiliation{University of Bonn, 53115 Bonn} 
  \author{J.~Dingfelder}\affiliation{University of Bonn, 53115 Bonn} 
  \author{I.~Adachi}\affiliation{High Energy Accelerator Research Organization (KEK), Tsukuba 305-0801} 
  \author{H.~Aihara}\affiliation{Department of Physics, University of Tokyo, Tokyo 113-0033} 
  \author{K.~Arinstein}\affiliation{Budker Institute of Nuclear Physics SB RAS and Novosibirsk State University, Novosibirsk 630090} 
  \author{D.~M.~Asner}\affiliation{Pacific Northwest National Laboratory, Richland, Washington 99352} 
  \author{T.~Aushev}\affiliation{Institute for Theoretical and Experimental Physics, Moscow 117218} 
  \author{A.~M.~Bakich}\affiliation{School of Physics, University of Sydney, NSW 2006} 
  \author{K.~Belous}\affiliation{Institute for High Energy Physics, Protvino 142281} 
% ...
\collaboration{The Belle Collaboration}
\noaffiliation

\maketitle

{\renewcommand{\thefootnote}{\fnsymbol{footnote}}}
\setcounter{footnote}{0}

\section{Introduction}

Semileptonic decays of $b$-flavored mesons constitute a very important class of decays
for determination of the elements of the Cabibbo-Kobayashi-Maskawa (CKM) matrix~\cite{CKM},
$V_{ub}$ and $V_{cb}$, and for understanding the origin of $CP$ violation in the
Standard Model (SM).  Although semileptonic $B^0$ and $B^+$ meson decays have been 
precisely measured by experiments running at the $\Upsilon(4S)$ resonance, and have been
well studied in theory, experimental information on the decay of the
$B_s^0$ meson is relatively limited. The interest in the physics of
the $B_s^0$ has intensified in recent years, motivated by studies of
the dilepton production asymmetry in $b \bar b$ production~\cite{Abazov:2010hv}. Semileptonic
$B_s^0$ decays are used as a normalization mode for various searches for new
physics at hadron colliders~\cite{Aaij:2011jp}, and in the future with the next
generation $B$ factories.  Semileptonic $B_s^0$ decays also provide an analogous
approach to studying the CKM matrix elements and testing theoretical
predictions, as meson decays that involve a spectator strange quark can be 
predicted more accurately than analogous decays with a spectator up or down quark.

An important expectation from heavy quark theory that is exploited in studies of
$B_s^0$ decays is the equality relation, based on $SU(3)$ symmetry, 
between the semileptonic decay widths \cite{Gronau2010,bigi2011}:
\begin{equation}
\Gamma_\text{SL}(B_s^0) = \Gamma_\text{SL}(B^+) = \Gamma_\text{SL}(B^0)\,.
\label{su3}
\end{equation}
The presence of the heavier spectator strange quark introduces, however, some
amount of $SU(3)$ symmetry breaking, as observed in decays of open charm mesons
\cite{charmSL}. Theoretical predictions based on heavy quark symmetry in Refs.
\cite{Gronau2010,bigi2011} find that Eq.~\ref{su3} should hold for 
$B_{(s)}$ decays to the percent level, which must be tested in experiment. 
The BaBar collaboration has determined the branching fraction 
$\mathcal{B}(B_s \to X \ell \nu) = [9.5^{+2.5}_{-2.0}(\text{stat})^{+1.1}_{-1.9}(\text{syst})] \%$
in a dataset obtained from an energy scan above the $\Upsilon(4S)$ resonance by 
measuring the inclusive yields of $\phi$ mesons and $\phi\ell$ pairs 
that are more abundant in $B_s^0$ decays \cite{babarbsxlnu2012}.
The semileptonic $B_s$ width has  been studied in part
by the D0 and LHCb collaborations, which measured the exclusive decay modes 
$B_s \to D_{2s}^* \ell \nu$ and $B_s \to D_{1s} \ell \nu$ \cite{Abazov:2007wg,Aaij:2011ju}.
In this paper, we report a measurement of the $B_s^0 \to X^- \ell^+ \nu_\ell$ branching 
fractions for $\ell = e$ and $\mu$ separately and their weighted average. 
The measurements are the most precise to date.

\section{Data sample, detector and simulation}
The data used in this analysis were collected with the Belle detector
at the KEKB asymmetric energy $e^+e^-$ collider \cite{KEKB}. The 
Belle detector is a large-solid-angle magnetic spectrometer that consists of
a silicon vertex detector (SVD), a 50-layer central drift chamber (CDC), an
array of aerogel threshold Cherenkov counters (ACC),  
a barrel-like arrangement of time-of-flight scintillation counters (TOF), and an
electromagnetic calorimeter (ECL) comprised of CsI(Tl) crystals located inside 
a superconducting solenoid coil that provides a 1.5~T magnetic field.  An iron
flux-return located outside of the coil is instrumented to detect $K_L^0$ mesons
and to identify muons (KLM).  The detector is described in detail
elsewhere~\cite{Belle}.

The results in this paper are based on a $\unit[121]{fb^{-1}}$ data sample
collected near the $\Upsilon(5S)$ resonance ($\sqrt{s} = \unit[10.87]{GeV}$),
which contains $(7.1 \pm 1.3) \times 10^6$ $B_s^{0(*)}\bar{B}_s^{0(*)}$ 
pairs \cite{nbs}. An additional $\unit[63]{fb^{-1}}$ data sample taken at 
$\sqrt{s} = \unit[10.52]{GeV}$, below the energy threshold for $b$-flavored 
meson production (off-resonance) is used to subtract background arising from 
the continuum $e^+e^- \to q \bar q$ process.

We use Monte Carlo (MC) techniques to separately simulate the production of
$B_{u,d}$ ($B^+$, $B^0$) and $B_s^0$ mesons at the $\Upsilon(5S)$
resonance. Events are generated with the EVTGEN event generator \cite{evtgen}, and
then processed through the detector simulation implemented in GEANT3 \cite{geant3}.   
The simulated samples of $B$-pair
events are equivalent to six times the integrated luminosity of the data.  
For the simulation of signal semileptonic $B_s^0$ decays, 
the lack of exclusive measurements of this system forces us to
rely on prior knowledge in the $B_{u,d}$ systems and employ a variety of phenomenological
models. First, we assume the composition of the $B_s^0$
semileptonic decay width is somewhat analogous to that of the $B^0$ 
system~\cite{Bailey:2012rr,Chen:2011ut,Li:2010bb,PDBook}. 
We include the following $B_s^0 \to X_c \ell \nu$ 
decay modes in the simulation, with their nominal branching fractions in parentheses: $X_c=$
$D_s$(2.1\%), $D_s^*$(4.9\%), $D_{s0}^*(2317)$(0.4\%), $D_{s1}(2460)$(0.4\%),
$D_{s1}(2536)$(0.7\%), and $D_{s2}^*(2573)$ (0.7\%). To simulate these decay modes,
we use the ISGW2 quark model \cite{ISGW2} for all modes, and an additional model based on
heavy quark effective theory (HQET) \cite{HQET2} for the $B_s \to D_s^{(*)} \ell
\nu$ modes. The form factors for the $B_s \to D_s^{(*)}
\ell \nu$ modes in the HQET parametrization are taken to be the same as in $B \to D^{(*)} \ell \nu$
decays, and the values taken from the Heavy Flavor Averaging Group \cite{hfag}. QED final
state radiation in semileptonic decays is added using the PHOTOS package
\cite{photos}.

\section{Measurement Overview}
Only one fifth of the mesons containing a $b$-quark
produced near the $\Upsilon(5S)$ resonance are $B_s^0$ mesons; the remainder are
$B_{u,d}$ mesons. In this analysis, the relative abundance of $B_s^0$ mesons 
is enhanced by reconstructing, or tagging, the CKM-favored $\bar{B}_s^0
\to D_s^+$ transition \cite{CC}, where $\mathcal{B}(B_s^0 \to D_s^\pm X) = (93
\pm 25)\%$ \cite{Bs2DsX}. The signal signature is a lepton ($e^+, \mu^+$) 
from the decay of the other $B_s^0$ in the event. 
To ensure that this lepton does not originate from the same $B_s^0$ meson 
as the reconstructed $D_s^+$ meson, $D_s^+\ell^+$ pairs are selected wherein 
the $D_s^+$ and $\ell^+$ have the same electric charge. 
The quantity obtained in the measurement is the ratio
\begin{equation}
\mathcal{R} = 
\frac{N_{D_s^+ \ell^+}}{N_{D_s^+}}~~\text{with}~\ell = e, \mu,
\label{eq:defratio}
\end{equation}
where $N_{D_s^+}$ and $N_{D_s^+\ell^+}$ are the efficiency-corrected yields of
$D_s^+$ and $D_s^+\ell^+$ pairs from $B_{(s)}$ decays. The ratio is proportional 
to the inclusive semileptonic branching fraction 
$\mathcal{B}(B_s^0 \to X^- \ell^+ \nu_\ell)$, plus dilution terms due to 
background $B_{u,d}$ decays. The yields from $B_{u,d}$ decays are approximately 
30\% and 15\% in the $D_s^+$ and $D_s^+\ell^+$ samples respectively, estimated 
using measured values of the $B_{u,d}$ and $B_s^0$ production fractions near the 
$\Upsilon(5S)$ resonance, their branching fractions to $D_s^+$ and 
$D_s^+\ell^+$ final states, and their mixing probabilities.

\section{Event selection}
\subsection{$D_s$ selection}
Charged particle tracks are required to originate from a region close to the
interaction point by applying the following selections on the impact parameters
along the $z$ axis (opposite the positron beam) and in the perpendicular
$r$-$\phi$ plane: $|dz|<2$ cm and $dr<0.5$ cm. In addition,
we demand at least one associated hit in the SVD detector. For pion and kaon
candidates, the Cherenkov light yield from the ACC, the time of 
flight information from the TOF, and the specific ionization $dE/dx$ from the 
CDC are required to be consistent with the appropriate mass hypotheses. 

Candidate $D_s^+$ mesons are reconstructed in the cleanest decay mode $D_s^+ \to
\phi \pi^+$, with the $\phi$ resonance reconstructed via $\phi \to K^+K^-$. 
The reconstructed $\phi$ and $D_s^+$ masses are required to lie within $\pm
\unit[8]{MeV}$ and $\pm \unit[65]{MeV}$ of the nominal $\phi$ and $D_s^+$
masses~\cite{PDBook}. The corresponding $\phi$ selection efficiency is $99\%$.
To suppress misreconstructed $D_s^+$ mesons, we require $|\cos \theta_{\rm h}| > 0.5$.
The helicity angle $\theta_{\rm h}$ is defined as the angle 
between the reconstructed $D_s^+$ momentum and the $K^-$ momentum in the $\phi$ 
rest frame. Non-resonant 
$D_s^+ \to KK\pi$ decays (such as from S-wave processes) passing the selection criteria
are treated as signal. Multiple $D_s^+$ candidates per event are allowed.
Correctly reconstructed $D_s^+$ mesons from continuum background are produced 
directly in processes of the type $e^+e^- \to c\bar{c}$ $\to D_s^\pm X$, and typically have 
high momenta  $p^*(D_s^+)$ in the center-of-mass (CM) frame of the $e^+e^-$ beams
with a maximum of $p^*_{\text{max}}(D_s^+) = \sqrt{s/4 - m(D_s^+)^2}$ \cite{speedoflight}. 
The maximum CM momentum of $D^+_s$ mesons produced in $B_s^0$ decays is half that of 
direct production, due to restricted decay phase space.
Therefore, to suppress events from the continuum background we require
\begin{equation}
x(D_s^+) = \frac{p^*(D_s^+)}{p^*_{\text{max}}(D_s^+)} =
\frac{p^*(D_s^+)}{\sqrt{s/4 - m(D_s^+)^2}} < 0.5\,.
\end{equation}

\subsection{Lepton selection }
Each $D_s^+$ candidate is combined with an electron or
muon having the same-sign charge. Electron candidates 
are identified using the ratio of the energy
detected in the ECL to the track momentum, the ECL shower shape, position
matching between the track and ECL cluster, the energy loss in the CDC, and the
response of the ACC counters. Muons are identified based on their penetration
range and transverse scattering in the KLM detector. The polar acceptance
regions are $18^\circ < \theta < 150^\circ$ and $25^\circ < \theta < 145^\circ$
for electrons and muons, respectively.  Leptons are reconstructed with a minimum
lepton momentum in the lab frame $p(\ell^+)$ of 0.6 GeV corresponding to the 
acceptance threshold of the detector. Lepton candidates are rejected if they are likely to have
originated from $J/\psi$ decays, using the mass criterion
$|m(\ell^+h^-) - m(J/\psi)| < \unit[5]{MeV}$, where $h^-$ is any charged track
with a mass hypothesis based on the signal candidate lepton.
Electrons that appear to originate from Dalitz $\pi^0$ decays and from
converted photons are removed by requiring
$|m(\ell^+h^-\gamma) - m(\pi^0)| < \unit[32]{MeV}$ and $|m(\ell^+h^-)| <
\unit[100]{MeV}$, 
respectively, where $h^-$ is defined as above and 
$\gamma$ is any detected photon. The lepton identification 
efficiencies multiplied by the geometrical acceptance
are 77\% (electrons) and 71\% (muons). The probabilities that a selected lepton 
candidate is a misidentified charged kaon or pion are 6\% and 19\% for electrons 
and muons, respectively.

The lepton detection efficiencies and misidentification probabilities in the MC 
simulation are calibrated to data. The calibration factors for the detection 
efficiencies are obtained from the study of
$\gamma\gamma \to \ell^+\ell^-$ and $J/\psi \to \ell^+\ell^-$.
The misidentification probabilities are determined from 
$D^{*+} \to D^0 \pi^+_\text{slow},~D^0 \to K^-\pi^+$ decays by studying the 
electron and muon likelihood of the $K^-$ and $\pi^+$ tracks 
from the $D^0$. The pion from the $D^{*+}$, $\pi^+_\text{slow}$, has a momentum of 
only a few hundred $\unit{MeV}$ as it is produced just above the kinematic threshold. 

\section{Fit results}

 \begin{figure*}
\includegraphics[width=0.8\textwidth]{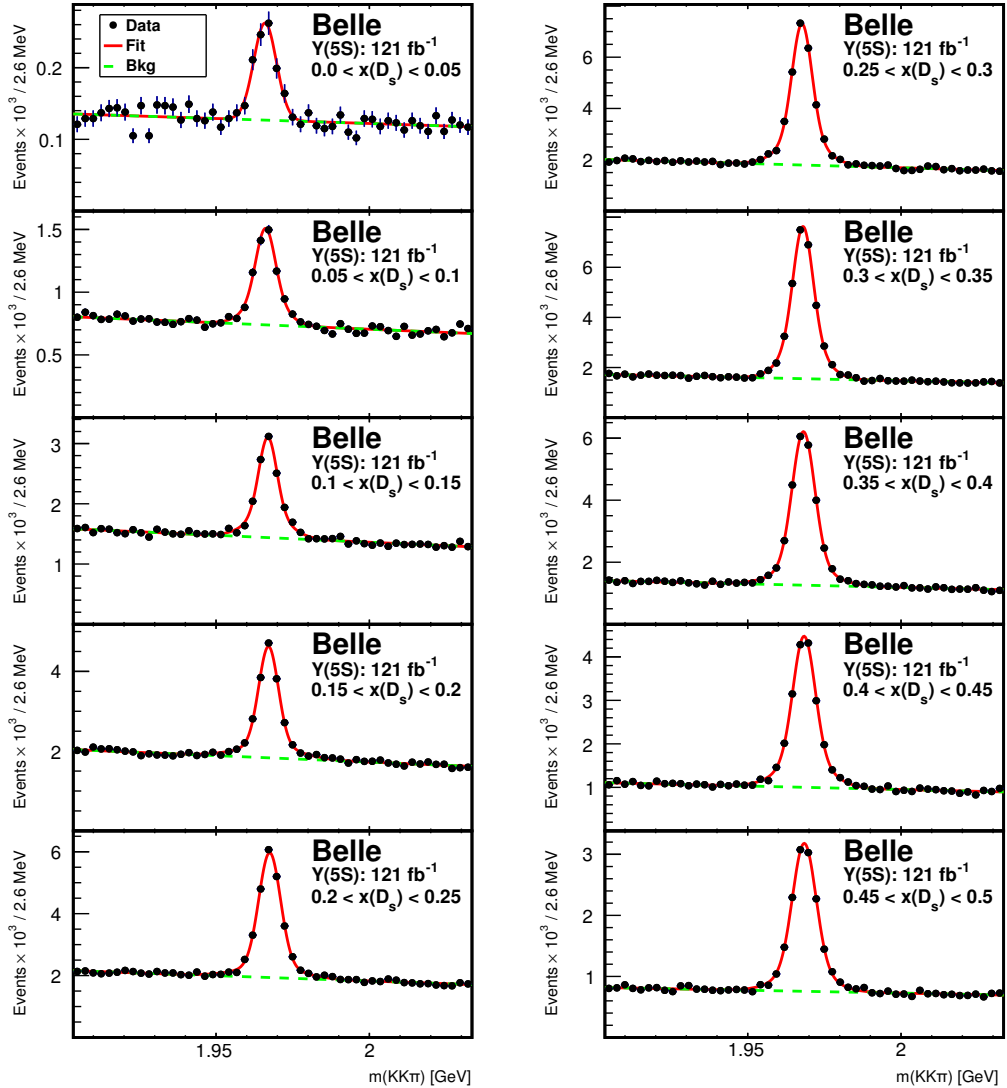}
\caption{The invariant $KK\pi$ mass spectra collected near the $\Upsilon(5S)$ resonance 
in bins of normalised $D_s^+$ momentum, $x(D_s^+)$, in the signal region 
($x(D_s^+) < 0.5$). The fits are used to determine the total number of $D_s^+$ mesons 
from $b$-flavored mesons in the $\Upsilon(5S)$ sample.} 
\label{fig:kkpimass_5s}
\end{figure*}

The number of $D_s^+$ mesons in data is determined from fits to the $KK\pi$ mass
distribution. The signal shape used in the fit is modeled as
two Gaussian functions with a common mean; the combinatorial background is
modeled by a linear function. The fit parameters are the
normalizations of signal ($N_\text{sig}$) and background ($N_\text{bkg}$), the slope of the linear
function ($b$) and the parameters of the two Gaussian functions: the common mean ($\mu_{\rm sig}$), the width of one
Gaussian ($\sigma_1$), the ratio of the widths ($r_\sigma = \sigma_2/\sigma_1$) and the ratio of the
normalizations ($r_N$). 
\begin{figure*}
\subfigure[]{
\includegraphics[width=0.3\textwidth]{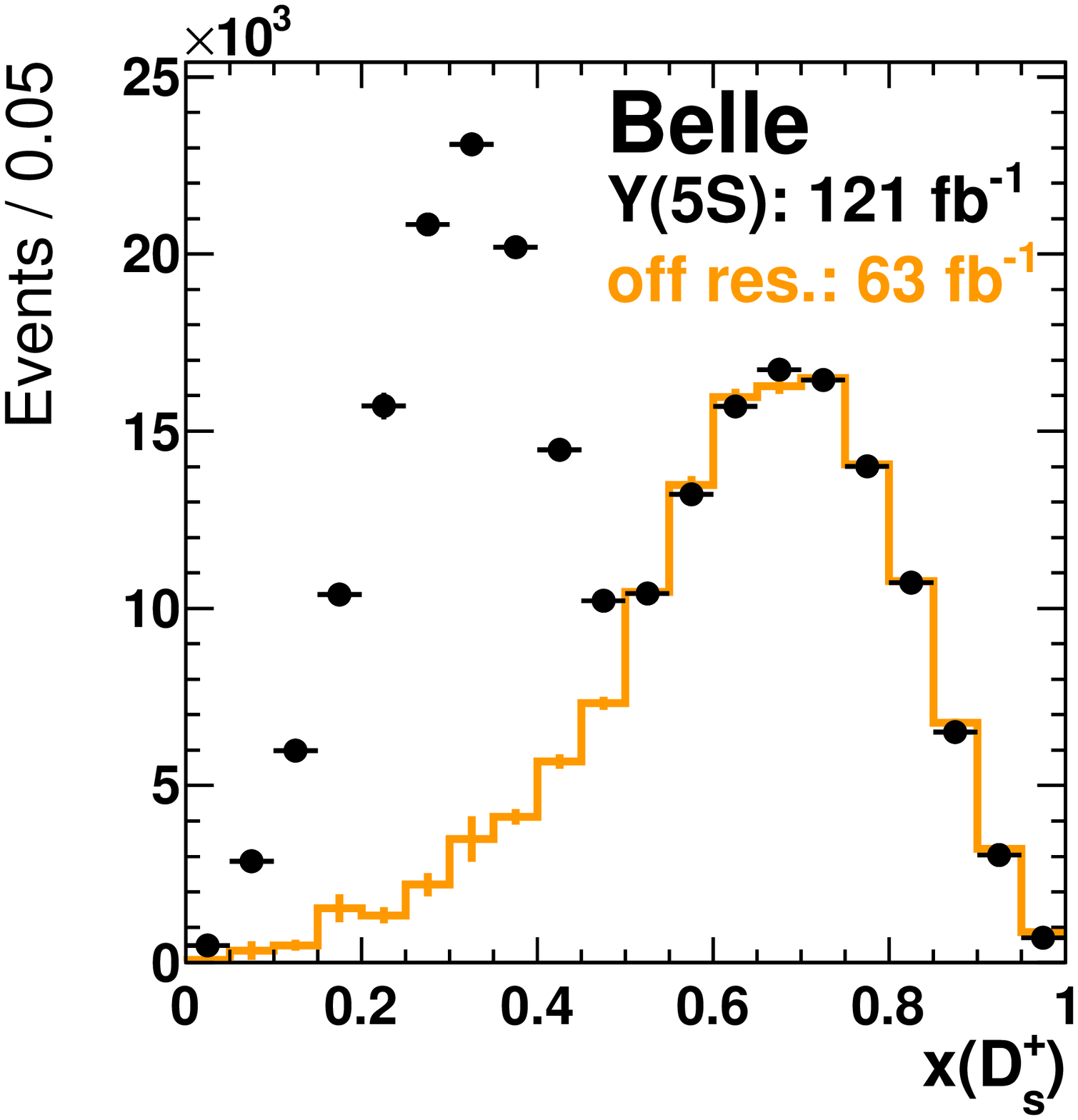}
\label{fig:xDs}
}
\subfigure[]{
\includegraphics[width=0.3\textwidth]{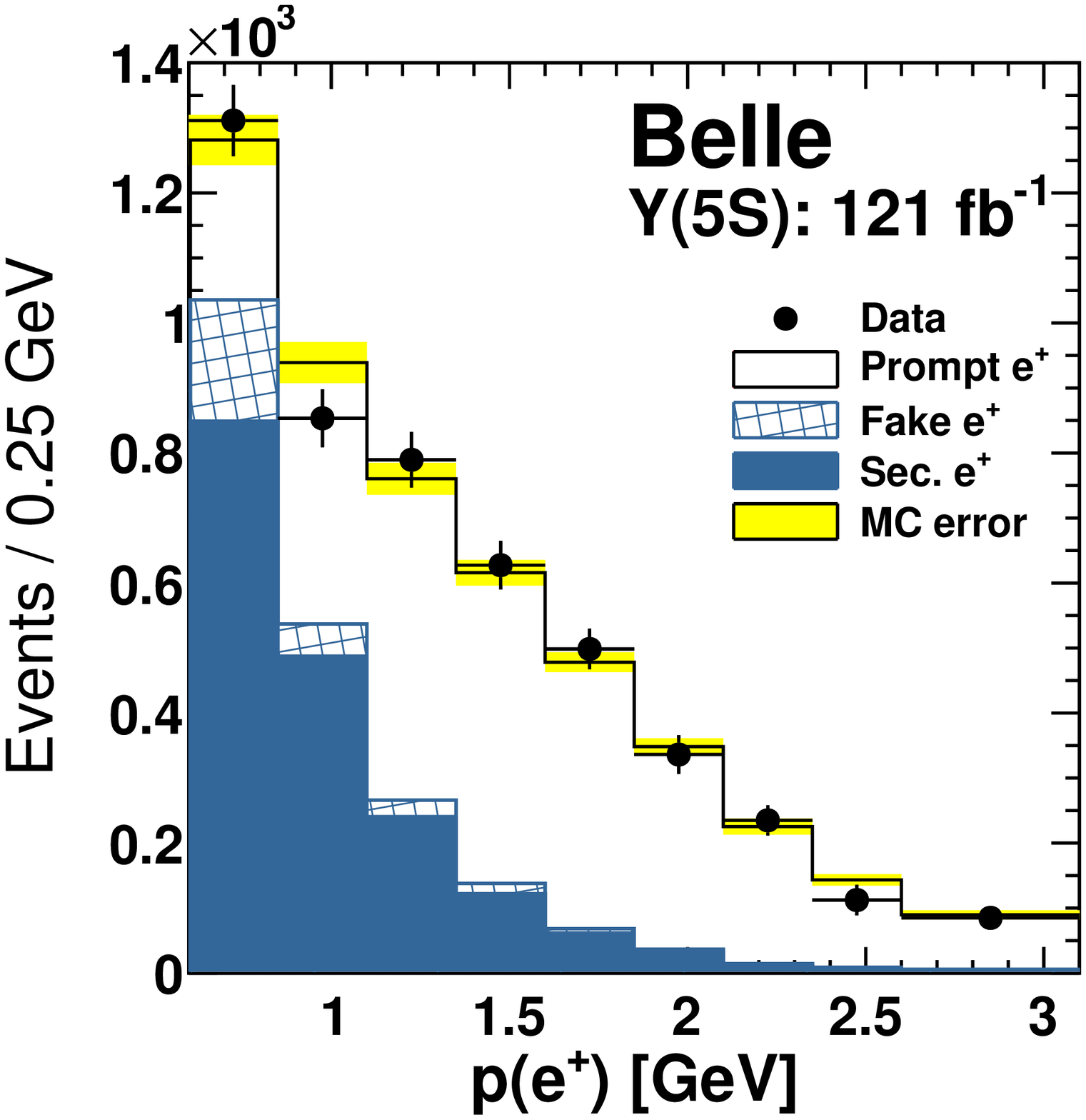}
\label{fig:elecmomfit}
}
\subfigure[]{
\includegraphics[width=0.3\textwidth]{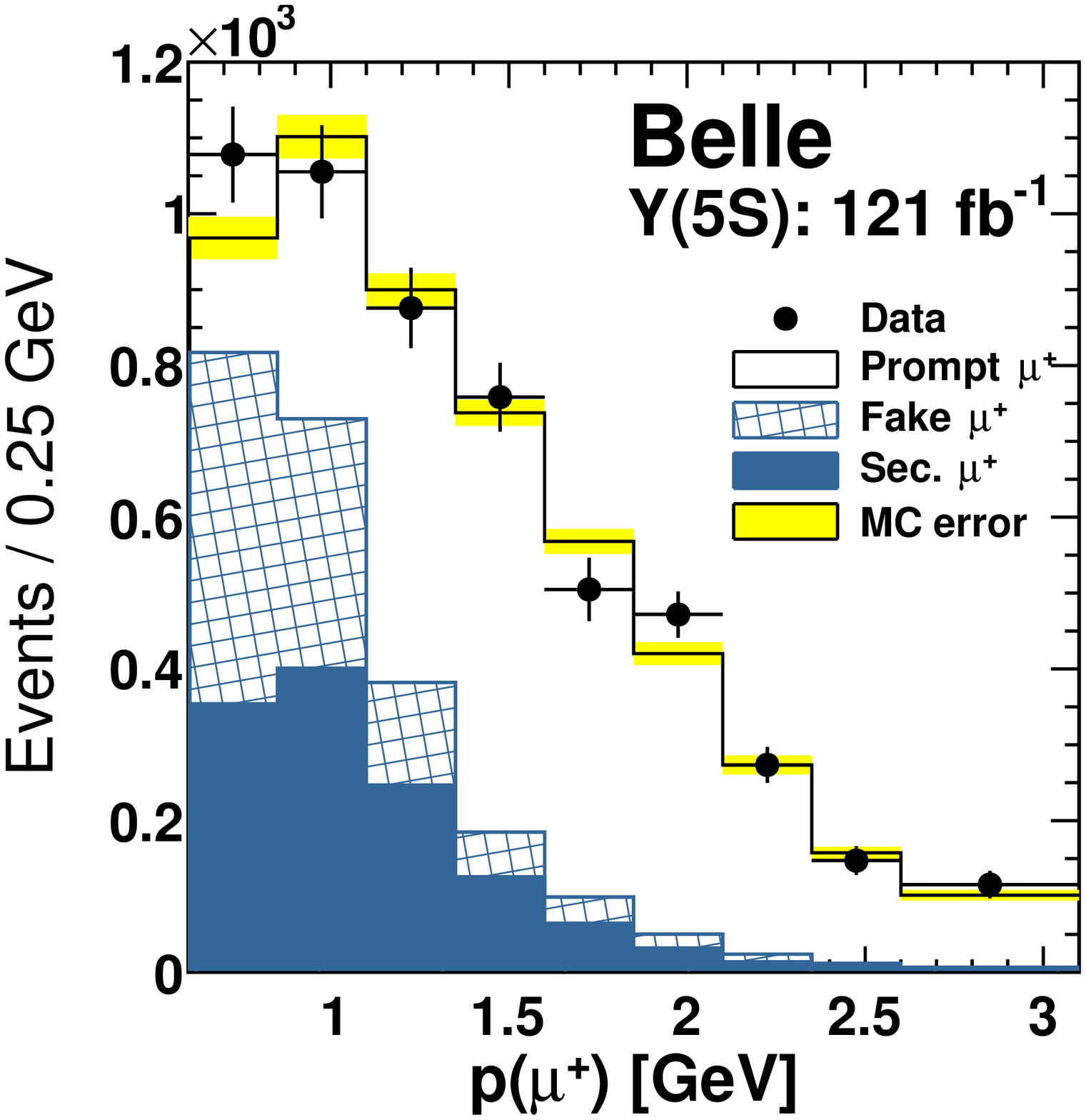}
\label{fig:muonmomfit}
}
\caption{Momentum spectra obtained from $KK\pi$ mass fits: (a) In bins of $x(D_s^+)$
($D_s^+$ sample); (b)+(c) In bins of $p(e^+)$ and $p(\mu^+)$, respectively, where continuum
backgrounds have been subtracted using off-resonance data ($D_s^+\ell^+$ sample).
The MC uncertainty (yellow) includes both statistical and systematic uncertainties.}
\end{figure*}

For the measurement of $N_{D_s^+}$, the fits to $m_{KK\pi}$ are performed in 20 equal bins of 
the normalized $D_s^+$ momentum $x(D_s^+)$ in the full range $[0, 1]$, including 
the control region $x(D_s^+)>0.5$. A binned approach is used to accommodate  $x(D_s^+)$ dependence on 
the signal and background shape parameters ($\mu_{\rm sig}$, $\sigma_1$, $b$). 
  
The fit results for the parameters $r_\sigma$ and $r_N$ are found to be independent of $x(D_s^+)$. 
Figure \ref{fig:kkpimass_5s} shows the $KK\pi$ mass fits for $\Upsilon(5S)$ data in the signal region 
($x(D_s^+) < 0.5$) and Fig. \ref{fig:xDs} the obtained $D_s^+$ momentum spectra for 
$\Upsilon(5S)$ data and off-resonance data. The off-resonance data is scaled with a factor 
$S_\text{cont} = (\mathcal{L}_{\Upsilon(5S)}/s_{\Upsilon(5S)})/(\mathcal{L}_{\text{off}}/s_{\text{off}})$
$= 1.81 \pm 0.02$ to account for the difference in integrated luminosities and the
dependence of the quark pair production cross section on the center-of-mass energy $\sqrt{s}$. 

The total $N_{D_s^+}$ is obtained by integrating over the region 
$x(D_s^+) < 0.5$ and subtracting the continuum background given by the scaled off-resonance 
distribution. A total of  $[12.42 \pm 0.08(\text{stat})] \times 10^4$ $D_s^+$ mesons are reconstructed, 
where $[2.7 \pm 0.1 (\text{stat})] \times 10^4$ of these are from continuum processes.
This approach is validated by taking the difference between $\Upsilon(5S)$ and 
off-resonance data in the control region $x(D_s^+)>0.5$, where only events from the continuum can 
contribute. The difference is found to be $-872 \pm 1778$, consistent with the expectation of zero.

\begin{figure*}
\includegraphics[width=0.8\textwidth]{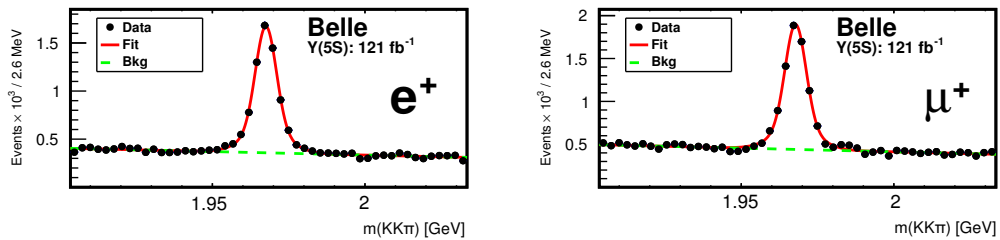}
\caption{The invariant $KK\pi$ mass spectra for the full
sample of selected $D_s^+\ell^+$ pair events, collected near the
$\Upsilon(5S)$ resonance. The figure shows the fits used to determine 
the shape parameters for the fits in bins of $p(\ell^+)$ 
(see text for details).
} 
\label{fig:kkpimassfit}
\end{figure*}

For the $N_{D_s^+ \ell^+}$ measurements, the $KK\pi$ mass fits are performed in
nine bins of lepton momentum in the range $\unit[0.6]{GeV} < p(\ell^+) < \unit[3.1]{GeV}$, where 
the lower and upper thresholds are chosen due to the detection sensitivity to electrons and muons, 
and to the semileptonic decay kinematic endpoint, respectively 
(see Figs.~\ref{fig:elecmomfit} and \ref{fig:muonmomfit}). The $D_s^+ \ell^+$
samples do not contain enough events to determine all seven fit parameters. 
Therefore $r_\sigma$ and $r_N$ are fixed to the values obtained in the $N_{D_s^+}$ measurement. 
The remaining parameters, other than $N_\text{sig}$ and $N_\text{bkg}$, are determined from a fit to 
the total $D_s^+\ell^+$ sample without the binning in $p(\ell^+)$ as shown in Fig.~\ref{fig:kkpimassfit}. The $\chi^2 / \text{ndf}$ 
of the fits over the full lepton momentum range are found to be $37 / 45$ and $58 / 45$ for electrons and muons, respectively.
Compared to the $D_s^+$ sample, the continuum background in the $D_s^+\ell^+$ sample is 
suppressed due to the same-sign lepton requirement. The remaining continuum background is 
subtracted using scaled off-resonance data. The shape difference of the continuum lepton 
momentum spectra at the $\Upsilon(5S)$ and in the off-resonance samples is determined from MC simulation and
the effect is corrected by a bin-by-bin re-weighting before the subtraction. 

A $\chi^2$ fit to the lepton momentum spectrum is performed with two components:
the prompt lepton signal and the remaining $B_{u,d,s}$ backgrounds, which is
the sum of secondary leptons (not coming directly from $B_{u,d,s}$ decays) and misidentified lepton
candidates. The shapes of the signal and $B_{u,d,s}$ backgrounds are derived from MC
simulation. Figures \ref{fig:elecmomfit} and \ref{fig:muonmomfit} show the fit result, and
the goodness of the fits is listed in Table \ref{tab:ratios}.  
The numbers of prompt leptons obtained in the fit are corrected for efficiency
and geometrical acceptance. The results are extrapolated from the experimental momentum 
threshold of $p(\ell^+)>0.6$ GeV to the full phase space region  
using MC simulation, where the uncertainty on this acceptance is included in the 
systematic uncertainties. The signal acceptance in the region $p(\ell^+)>\unit[0.6]{GeV}$ is $93\%$ for electrons and
$94\%$ for muons. Finally, we find $[3.91\pm 0.18(\text{stat})] \times 10^3$ 
and $[4.37 \pm 0.21(\text{stat})] \times 10^3$ prompt signal electrons and muons, respectively. 
To determine ${\cal R}$ (Eq. \ref{eq:defratio}), we additionally account for 
the difference in $D_s^+$ reconstruction efficiencies
between the inclusive $D_s^+$ and the signal samples $D_s^+\ell^+$. These efficiencies
take into account the possibility of more than one $D_s^+ \to \phi(K^+K^-)\pi^+$ decay
per event. The $D_s^+$ reconstruction efficiencies and the results for ${\cal R}$ are summarized in
Table \ref{tab:ratios}, where the combined result is obtained from the weighted
average of the $e^+$ and $\mu^+$ modes, taking into account measurement
correlations.

\begin{table}
\caption{Measured ratios $\mathcal{R}$. The first uncertainty is statistical; 
the second is systematic. The last row shows the result for the combination of the $e^+$ and $\mu^+$ modes
and takes into account the correlations.}
\begin{tabular}{lcccc}
\hline \hline
Mode & Ratio $\mathcal{R} \times 10^{-4}$ & $\chi^2/\text{ndf}$ &
$\epsilon_{D_s^+}(KK\pi)$ & $\epsilon_{D_s^+\ell^+}(KK\pi)$ \\
\hline
$e$ & $394 \pm 20 \pm 13$ & 3.6 / 7 & $28.2 \%$ & $28.7\%$\\
$\mu$ & $432 \pm 22 \pm 17$ & 5.7 / 7 & $28.2 \%$ & $29.2\%$\\
$e,\mu$ & $409 \pm 15 \pm 14$ & --- & --- & --- \\
\hline \hline
\end{tabular}
\label{tab:ratios}
\end{table}

\section{Systematic uncertainties on ${\cal R}$}

The systematic uncertainties on the ratio $\mathcal{R}$ are divided into four 
categories: detector effects, fitting procedure, background modeling and
signal modeling. They are discussed in turn below, and are given as relative
uncertainties. They are also summarized in Tab. \ref{tab:allsystematics}.

Numerous potential systematic uncertainties that relate to the reconstruction
of the $D_s^+$ ultimately cancel in the ratio; these include uncertainties associated
with kaon and pion reconstruction. The uncertainty on the calibration of
the electron (muon) identification is 0.7\% (1.4\%).
The uncertainty on the lepton misidentification is below 0.1\%. Another 0.4\% uncertainty is added
for the reconstruction efficiency of the lepton track. The statistical uncertainty of 
the efficiencies $\epsilon_{D_s^+ e^+}(KK\pi)$ and $\epsilon_{D_s^+ \mu^+}(KK\pi)$ is 0.8\%.

Uncertainties in the modeling of the $KK\pi$ mass shape cancel in the ratio
$\mathcal{R}$. The shape parameters fixed in the $N_{D_s^+ \ell^+}$ fits are 
each varied by one standard deviation and the variations on the fit results are
added in quadrature to determine the systematic uncertainty.
This results in an uncertainty of 2.0\% (2.2\%) for electrons (muons).

The scale factor $S_\text{cont}$ for the off-resonance data and the correction
of the off-resonance lepton momentum spectrum add uncertainties of 0.4\%
and 1\%, respectively. The knowledge of the composition of the fit component 
containing the combined background from secondary leptons and and misidentified lepton 
candidates is limited by the precision of the measurements of $B_s^0$ branching fractions, which
is estimated to be of the order of $30\%$. Hence, the yields of secondary leptons from
$D_{u,d,s}$, from $\tau$ and from other decays, as well as the rate of misidentified
leptons are scaled by $\pm30\%$ and the variation of $\mathcal{R}$ is taken as
systematic uncertainty, giving 1.0\% (1.5\%) for electrons (muons). 

For the signal model, since most of the exclusive modes have not been measured,
the shape uncertainty is estimated as the full difference between the result
obtained with HQET and with the ISGW2 model where applicable. For both electrons and muons,
the obtained uncertainty is 1.0\%. Since the background from $B_{u,d}$ decays is expected
to be approximately $15 \%$ of the measured semileptonic yield and the 
semileptonic width of $B_{u,d}$ decays has been studied in more detail, the shape 
uncertainties are found to be negligible compared to $B_s^0$ decays.

The uncertainty due to the composition of the semileptonic width is
evaluated by varying the normalization of each mode by $\pm30\%$ and adding the
uncertainties in quadrature. The resulting uncertainties on ${\cal R}$ are
1.0\% and 1.1\% for electrons and muons, respectively. Due to the inclusiveness of 
the analysis, the total uncertainty on the signal lepton acceptance is only $0.3\%$.

The total systematic uncertainty on $\mathcal{R}$ is calculated by summing the
above uncertainties in quadrature. It is found to be 3.3\% (3.1\%) for electrons and 
4.0\% (3.6\%) for muons, where
the values in parentheses are the fully correlated errors between both modes.
Taking these correlations into account, the total systematic uncertainty on the
combined value of ${\cal R}$ is 3.0\%.

\begin{table}
\caption{Overview of the relative systematic uncertainties of the ratio $\mathcal{R}$.}
\label{tab:allsystematics}
\begin{tabular}{lrr}
\hline \hline
Uncertainty [\%] & $e$ & $\mu$ \\
\hline 
{\bf Detector effects} & & \\
Lepton identification & 0.7 & 1.4 \\
Fake lepton rate & $<0.1$ & $<0.1$  \\
Tracking efficiency & 0.4 & 0.4 \\
$D_s^+$ reconstruction efficiencies & 0.8 & 0.8 \\
\hline
{\bf Fitting procedure} & & \\
Shape error in $KK\pi$ mass fits& 2.0 & 2.2 \\
\hline
{\bf Background modeling} & & \\
Continuum scale factor $S_\text{cont}$ & 0.4 & 0.4 \\
Kinematic smearing of $p(\ell)$ from continuum & 1.0 & 1.0  \\
Secondary and and fake $\ell$ bkg. composition & 1.0 & 1.5 \\
\hline
{\bf Signal modeling} & & \\
Shape of the prompt lepton spectrum & 1.0 & 1.0 \\
Composition of the semileptonic width & 1.6 & 2.1 \\
\hline
{\bf Total} & {\bf 3.3} & {\bf 4.0} \\ 
Total correlated & 3.1 & 3.6 \\
\hline \hline
\end{tabular}
\end{table}

\section{Extraction of the branching fraction}

The extraction of the $B_s^0 \to X^- \ell^+ \nu_\ell$ branching fraction is
based on a prediction of the measured ratio $\mathcal{R}$ and includes the estimation
of the background from $B_{u,d}$ decays. 
This approach is based on the calculation of the number of 
same-sign lepton pairs $\ell^+\ell^+$ in $\Upsilon(5S)$ decays discussed in 
Refs. \cite{phdlouvot,Sia:2006cq}.
The measured yields $N_{\zeta}$ (where $\zeta = D_s^+,~D_s^+\ell^+$) 
contain a contribution from $B_s^0$ decays 
$\mathcal{N}_{\zeta}(B_s^{(*)}\bar{B}_s^{(*)})$ and from $B_{u,d}$ background 
$\mathcal{N}_{\zeta}(B_{u,d}^{(*)}\bar{B}_{u,d}^{(*)}(\pi))$:
\begin{equation}
\mathcal{R} =
\frac{\mathcal{N}_{D_s^+\ell^+}(B_s^{(*)}\bar{B}_s^{(*)}) +
\mathcal{N}_{D_s^+\ell^+}(B_{u,d}^{(*)}\bar{B}_{u,d}^{(*)}(\pi))}
{\mathcal{N}_{D_s^+}(B_s^{(*)}\bar{B}_s^{(*)}) +
\mathcal{N}_{D_s^+}(B_{u,d}^{(*)}\bar{B}_{u,d}^{(*)}(\pi))} \,.
\label{eq:ratioeqfrac}
\end{equation}

The total number of produced $b$-quark pairs, $N_{b\bar{b}}$, cancels 
in the ratio.
Pairs of $b\bar{b}$ quarks produced near the $\Upsilon(5S)$ resonance hadronize in 
pairs of $B_{u,d}$ mesons with a probability of
$f_{ud} = f_u + f_d$, where
$f_u = \mathcal{B}(\Upsilon(5S) \to B^\pm X)/2 = (36.1 \pm 3.2) \%$ and 
$f_d = \mathcal{B}(\Upsilon(5S) \to B^0 X)/2 = (38.5 \pm 4.2)\%$
\cite{Drutskoy:2010an}.
$B_s^0$ pairs are formed with a probability of $f_s = (19.9 \pm 3.0)\%$ \cite{PDBook}. 
The remaining contribution to the $Y(5S) \to b\bar{b}$ decay width are 
bottomonium resonances, but no subsequent decays of these resonances to 
$D_s^+$ mesons have been observed so far. The contribution from
bottomonium is assumed to be negligible in the ratio $\mathcal{R}$, and
neglected in the calculations.

The production of $B_{u,d}$ mesons near the $\Upsilon(5S)$ center-of-mass
energy is divided in three classes \cite{Drutskoy:2010an}: two-body decays
$B_{u,d}^{(*)}\bar{B}_{ud}^{(*)}$,
three-body decays with an additional pion $B_{u,d}^{(*)}\bar{B}_{ud}^{(*)}\pi$ and 
the initial state radiation (ISR) process 
$e^+e^- \to \gamma_\text{ISR} \Upsilon(4S) \to \gamma_\text{ISR} B_{u,d}\bar{B}_{ud}$.
The fractions of the different two-body production mechanisms  are given by
the parameters $F_{B\bar{B}}$, $F_{B^*\bar{B}}$ and $F_{B^*\bar{B}^*}$, and
their sum is denoted by $F_2$. The fraction of three-body decays
is $(f_{ud}-F_2) \cdot F^\prime_3$, where $F^\prime_3 = F^\prime_{B\bar{B}\pi} + 
F^\prime_{B^*\bar{B}\pi} + F^\prime_{B^*\bar{B}^*\pi}$. 
The remainder $(f_{ud}-F_2)\cdot(1-F^\prime_3)$ is attributed to the ISR process.
From isospin symmetry, one can deduce that one-third of the
three-body decay modes are $B^{+(*)}B^{-(*)}\pi^0$ and $B^{0(*)}\bar{B}^{0(*)}\pi^0$,
with the remainder being $B^{+(*)}\bar{B}^{0(*)}\pi^-$ or $B^{-(*)}B^{0(*)}\pi^+$.

The mixing probability $\chi_q^{(\mathcal{C})}$ of a pair of $B_q^0$ mesons ($q=d,s$)
depends on $x_q = \Delta m_{B_q^0} / \Gamma_{B_q^0}$ and the $\mathcal{C}$ eigenstate 
in which the pair is produced:
\begin{equation}
\chi_q^{(+)} = \frac{x_q^2(3+x_q^2)}{2(1+x_q^2)^2}\quad \text{and} \quad      
\chi_q^{(-)} = \frac{x_q^2}{2(1+x_q^2)}\,.
\end{equation}
In contrast to $B^0$ mesons, where $x_d = 0.770 \pm 0.008$ 
\cite{PDBook}, $x_s = 26.49 \pm 0.29$ \cite{PDBook} is so large for $B_s^0$  mesons 
that the difference between even and odd $\mathcal{C}$ eigenstates can be
neglected. We use the approximation $\chi_s = (1-\chi_s) = 0.500 \pm 0.001$.
For $B^0$ produced together with a charged $B^-$ meson, the mixing probability is 
the same as for $\mathcal{C}=-1$. With this information, the contributions $\mathcal{N}$
to the yields from each $b$-flavored meson production mode can be calculated. The
factor of two takes into account the possibility that the reconstructed $D_s^+$ meson can stem from 
either of the two $b$-flavored mesons.
\begin{widetext}
\begin{equation}
\begin{array}{l}
\mathcal{N}_{D_s^+}(B_s^{(*)}\bar{B}_s^{(*)}) / N_{b\bar{b}} = 2 \cdot f_s \cdot \mathcal{B}(B_s^0 \to D_s^\pm X)\,,
\end{array}
\end{equation}
\begin{equation}
\begin{array}{l}
\mathcal{N}_{D_s^+}(B_{u,d}^{(*)}\bar{B}_{u,d}^{(*)}(\pi)) / N_{b\bar{b}} = 2 \cdot f_d \cdot \mathcal{B}(B^0 \to D_s^\pm X) + 2 \cdot f_u \cdot \mathcal{B}(B^+ \to D_s^\pm X)\,, 
\end{array}
\end{equation}
\begin{equation}
\begin{array}{l}
\mathcal{N}_{D_s^+ \ell^+}(B_s^{0(*)}\bar{B}_s^{0(*)}) / N_{b\bar{b}} =  2 \cdot f_s \cdot \mathcal{B}(B_s^0 \to X^- \ell^+ \nu_\ell)  \cdot (1 - \chi_s) \cdot \mathcal{B}(B_s^0 \to D_s^\pm X)\,,
\end{array}
\end{equation}
\begin{equation}
\begin{array}{llll}
\multicolumn{4}{l}{\mathcal{N}_{D_s^+\ell^+}(B_{u,d}^{(*)}\bar{B}_{u,d}^{(*)}(\pi))/N_{b\bar{b}} =} \\ 
&& 2 \cdot \frac{f_d}{f_{ud}} \cdot & \left[ F_{B\bar{B}} + F_{B^*\bar{B}^*} + \frac{1}{3}(f_{ud} - F_2) \cdot (F^\prime_{B\bar{B}\pi} + F^\prime_{B^*\bar{B}^*\pi}) + (f_{ud} - F_2) \cdot (1 - F_3^\prime) \right] \cdot \\
& && \underbracket[0.5pt]{\left\{\chi_d^{(-)} \cdot \mathcal{B}(B^0 \to D_s^+ X) + \left(1 - \chi_d^{(-)}\right) \cdot \mathcal{B}(B^0 \to D_s^- X) \right\} \cdot \mathcal{B}(B^0 \to X^- \ell^+ \nu_\ell)}_{B^{0(*)}\bar{B}^{0(*)}~\text{pairs,}~\mathcal{C}~\text{even}} \\
& + &
2\cdot \frac{f_d}{f_{ud}} \cdot & \left[ F_{B^*\bar{B}} + \frac{1}{3}(f_{ud} - F_2) \cdot F^\prime_{B^*\bar{B}\pi}) \right] \cdot \\
& && \underbracket[0.5pt]{\left\{\chi_d^{(+)} \cdot \mathcal{B}(B^0 \to D_s^+ X) + \left(1 - \chi_d^{(+)}\right) \cdot \mathcal{B}(B^0 \to D_s^- X)\right\} \cdot \mathcal{B}(B^0 \to X^- \ell^+ \nu_\ell)}_{B^0\bar{B}^{0*}~\text{pairs,}~\mathcal{C}~\text{odd}}  \\
& + &2 \cdot \frac{f_u}{f_{ud}} \cdot & \underbracket[0.5pt]{\left[F_2 + \frac{1}{3}(f_{ud} - F_2) \cdot F^\prime_3 + (f_{ud} - F_2) \cdot (1 - F_3^\prime)) \right] \cdot \mathcal{B}(B^+ \to D_s^- X) \cdot \mathcal{B}(B^+ \to X \ell^+ \nu_\ell)}_{B^{+(*)}B^{-(*)}~\text{pairs}} \\
& + &&\left[\frac{2}{3} \cdot (f_{ud} - F_2) \cdot F_3^\prime \right] \cdot \\
& && \left( \left\{\chi_d^{(-)} \cdot \mathcal{B}(B^0 \to D_s^+ X) + \left(1 - \chi_d^{(-)}\right) \cdot \mathcal{B}(B^0 \to D_s^- X) \right\} \cdot \mathcal{B}(B^+ \to X \ell^+ \nu_\ell) + \right. \\
& && \underbracket[0.5pt]{\left. \left\{\chi_d^{(-)} \cdot \mathcal{B}(B^+ \to D_s^+ X) + \left(1 - \chi_d^{(-)}\right) \cdot \mathcal{B}(B^+ \to D_s^- X) \right\} \cdot \mathcal{B}(B^0 \to X^- \ell^+ \nu_\ell) \right)}_{B^{+(*)}\bar{B}^{0(*)}~\text{and}~B^{-(*)}B^{0(*)}~\text{pairs}}\,.
\end{array}
\end{equation}
Equation \ref{eq:ratioeqfrac} is solved for $\mathcal{B}(B_s^0 \to X^- \ell^+ \nu_\ell)$, which is the only unknown quantity.
\begin{equation}
\mathcal{B}(B_s^0 \to X^- \ell^+ \nu_\ell) = \frac{\left[\mathcal{N}_{D_s^+}(B_s^{(*)}\bar{B}_s^{(*)}) + \mathcal{N}_{D_s^+}(B_{u,d}^{(*)}\bar{B}_{u,d}^{(*)}(\pi))\right] \cdot \mathcal{R} - \mathcal{N}_{D_s^+\ell^+}(B_{u,d}^{(*)}\bar{B}_{u,d}^{(*)}(\pi))}{2 \cdot f_s \cdot (1-\chi_s) \cdot \mathcal{B}(B_s^0 \to D_s^\pm X) \cdot N_{b\bar{b}}}
\end{equation}
\end{widetext} 
The parameters used to calculate the $\mathcal{N}$ terms are summarized in Table 
\ref{tab:chi2systematics}. The uncertainties on $\mathcal{B}(B_s^0 \to X^- \ell^+ \nu_\ell)$ from the
external parameters are obtained by varying each of them in turn by their uncertainties;
for asymmetric uncertainties, the larger one is used. The external parameters
are treated as if they were uncorrelated. The correlations between the ratio
$\mathcal{R}$ and the external parameters measured at Belle are negligible.

\begin{table}
\centering
\caption{Central values used for the extraction of the branching fraction
$\mathcal{B}$. The relative systematic uncertainty $|\Delta \mathcal{B} / \mathcal{B}|$ 
is given for the combined measurement. Parameter values are taken from Ref. \cite{PDBook} unless
otherwise stated.}
\begin{tabular}{llc}
\hline\hline
Parameter &  Value  & $|\Delta \mathcal{B} / \mathcal{B}| [\%]$
   \\ 
\hline
$f_u = \mathcal{B}(\Upsilon(5S) \to B^\pm X)/2$  & $(36.1 \pm 3.2) \%$ \cite{Drutskoy:2010an} & $0.8$ \\
$f_d = \mathcal{B}(\Upsilon(5S) \to B^0 X)/2$  &  $(38.5 \pm 4.2)\%$ \cite{Drutskoy:2010an}& $0.5$ \\
$f_s$ & $(19.9 \pm 3.0)\%$ & $ 2.2$ \\ 
$\mathcal{B}(B_s \rightarrow D_s^\pm X)$ & $(93 \pm 25)\%$ \cite{Bs2DsX} & $ 4.0$ \\ 
$\mathcal{B}(B^{+} \rightarrow D_s^{+} X)$ & $(7.9 \pm 1.4)\%$  & $ 2.2$ \\ 
$\mathcal{B}(B^{0} \rightarrow D_s^{+}X)$ & $(10.3 \pm 2.1)\%$  & $ 1.5$ \\ 
$\mathcal{B}(B^{0} \rightarrow D_s^{-}X)$ & $(1.5 \pm 0.8)\%$ \cite{BaBarB2Ds}  & $ 1.4$ \\ 
$\mathcal{B}(B^{+} \rightarrow D_s^{-} X)$ & $(1.1 \pm 0.4)\%$ & $1.1$ \\ 
$\mathcal{B}(B^{0} \rightarrow X \ell^+ \nu_\ell )$ & $(10.33 \pm 0.28)\%$ & $0.4$ \\ 
$\mathcal{B}(B^{+} \rightarrow X \ell^+ \nu_\ell)$ & $(10.99 \pm 0.28)\%$   & $0.1$ \\ 
$F_{B^{\ast}\bar{B}^{\ast}}$ & $(38.1 \pm 3.4)\%$   & $ 0.1$ \\ 
$F_{B^{\ast}\bar{B}}$ & $(13.7 \pm 1.6)\%$   & $ 0.1$ \\ 
$F_{B\bar{B}}$ & $(5.5 \pm 1.6)\%$ &   $ 0.0$ \\ 
$F^{\prime}_{B^{\ast}\bar{B}^{\ast}\pi}$ & $(5.9 \pm 7.8)\%$ \cite{Drutskoy:2010an} & $ 0.1$ \\ 
$F^{\prime}_{B^{\ast}\bar{B}\pi}$ & $(41.6 \pm 12.1)\%$  \cite{Drutskoy:2010an} & $ 0.2$ \\ 
$F^{\prime}_{B\bar{B}\pi}$ & $(0.2 \pm 6.8)\%$  \cite{Drutskoy:2010an} & $ 0.0$ \\ 
$x_d$ & $0.771 \pm 0.008$  & $ 0.1$ \\ 
$\chi_s$ & $0.500 \pm 0.001$  & $ 0.2$ \\ 
\hline\hline
\end{tabular}
\label{tab:chi2systematics}
\end{table}

\section{Results and Discussion}

We obtain the following values for the semileptonic
branching fraction $\mathcal{B}(B_s^0 \to X^- \ell^+ \nu_\ell)$:
\begin{eqnarray}
\nonumber \ell = e~~ &[9.1\:\pm\:0.5(\text{stat})\:\pm\:0.6(\text{syst})]\%\,,\\
\nonumber \ell = \mu~~ &[10.2\:\pm\:0.6(\text{stat})\:\pm\:0.8(\text{syst})]\%\,,\\
\nonumber \ell = e, \mu~~ &[9.6\:\pm\:0.4(\text{stat})\:\pm\:0.7(\text{syst})]\%\,.
\end{eqnarray}
The last branching fraction is the combination of the electron and muon
mode measurements.
Our result is consistent with the measurement in Ref. \cite{babarbsxlnu2012}
and substantially improves on both the statistical and systematic precision.

Table \ref{tab:brafrasyst} summarizes the uncertainties of the branching
fractions. The dominant uncertainty arises from the external parameters.
This is typical for almost any $B_s^0$ absolute branching fraction
measurement where the $B_s^0$ production rate near the $\Upsilon(5S)$ resonance has to be estimated.
In this measurement, the critical parameters $f_s$ and $\mathcal{B}(B_s \to D_s^\pm X)$ appear in
the numerator and denominator of the ratio $\mathcal{R}$ and therefore
the respective uncertainties partially cancel. 
The measurement of the ratio $\mathcal{R}$ is kept independent of the extraction of
$\mathcal{B}(B^0 \to X^-\ell^+\nu_\ell)$, in order to facilitate the update
of the branching fraction when the precision of external measurements improves.

\begin{table}
\centering
\caption{Relative uncertainties on the branching fraction $\mathcal{B}(B_s^0 \to X \ell^+ \nu_\ell)$ 
 in percent, for the electron and muon mode, and their combination.}
\begin{tabular}{lp{0.7cm}p{0.7cm}p{0.8cm}}
\hline \hline
Uncertainty~$[\%]$& $e$ & $\mu$ & $e,\mu$ \\
\hline
Detector effects & 1.3 & 1.9 & 1.2 \\
Fitting procedure & 2.4 & 2.6  & 2.4 \\
Background modelling & 1.8 & 2.2 & 1.8 \\
Signal modelling & 2.1 & 2.8 & 2.4 \\ 
External parameters (see Tab. \ref{tab:chi2systematics}) & 5.6 & 5.9 & 5.6 \\
\hline
Total systematic & 6.8 & 7.5 & 6.9 \\
\hline
Statistical & 5.7 & 6.0 & 4.2 \\
\hline \hline 
\end{tabular}
\label{tab:brafrasyst}
\end{table}

Using the well measured lifetimes of the $B_s^0$ and $B^0$ mesons, and
$\mathcal{B}(B^0 \to X^-\ell^+\nu_\ell)$ \cite{PDBook}, 
the inclusive semileptonic width of the $B_s^0$ is determined to be
$\Gamma_\text{SL}(B_s^0) = (0.94 \pm 0.08) \cdot \Gamma_\text{SL}(B^0)$\,
which is consistent with the theoretical expectation~\cite{Gronau2010, bigi2011}.
This level of precision is already an important test of the theoretical description of 
semileptonic $B_s^0$ decays.  To fully understand $SU(3)$ symmetry breaking 
effects, the heavy quark parameters of semileptonic $B_s^0$ decays must be 
measured directly. This can be achieved through the analysis of spectral moments, 
although it will require full reconstruction techniques only feasible at a 
next generation flavor factory.

\section{Summary}
We measured the inclusive semileptonic $B_s^0$ branching 
fraction $\mathcal{B}(B_s^0 \to X^- \ell^+ \nu_\ell) =
[9.6 \pm 0.4(\text{stat}) \pm 0.7(\text{syst})]\%$.
This is the most precise measurement to date and in agreement
with the previous measurement \cite{babarbsxlnu2012} 
and theoretical expectations \cite{Gronau2010, bigi2011}. 

\section{Acknowledgements}
We thank the KEKB group for the excellent operation of the
accelerator; the KEK cryogenics group for the efficient
operation of the solenoid; and the KEK computer group,
the National Institute of Informatics, and the 
PNNL/EMSL computing group for valuable computing
and SINET4 network support.  We acknowledge support from
the Ministry of Education, Culture, Sports, Science, and
Technology (MEXT) of Japan, the Japan Society for the 
Promotion of Science (JSPS), and the Tau-Lepton Physics 
Research Center of Nagoya University; 
the Australian Research Council and the Australian 
Department of Industry, Innovation, Science and Research;
the National Natural Science Foundation of China under
contract No.~10575109, 10775142, 10875115 and 10825524; 
the Ministry of Education, Youth and Sports of the Czech 
Republic under contract No.~LA10033 and MSM0021620859;
the Department of Science and Technology of India; 
the Istituto Nazionale di Fisica Nucleare of Italy; 
the BK21 and WCU program of the Ministry Education Science and
Technology, National Research Foundation of Korea,
and GSDC of the Korea Institute of Science and Technology Information;
the Polish Ministry of Science and Higher Education;
the Ministry of Education and Science of the Russian
Federation and the Russian Federal Agency for Atomic Energy;
the Slovenian Research Agency;  the Swiss
National Science Foundation; the National Science Council
and the Ministry of Education of Taiwan; and the U.S.\
Department of Energy and the National Science Foundation.
This work is supported by a Grant-in-Aid from MEXT for 
Science Research in a Priority Area (``New Development of 
Flavor Physics''), and from JSPS for Creative Scientific 
Research (``Evolution of Tau-lepton Physics'').


\begin{thebibliography}{99}
\bibitem{CKM}
N. Cabibbo, Phys. Rev. Lett. 10, 531 (1963); 
M. Kobayashi and T. Maskawa, Prog. Theor. Phys. 49, 652 (1973).

\bibitem{Abazov:2010hv}
V.~Abazov {\it et al.}  [D0 Collaboration],
Phys.\ Rev.\ D {\bf 82}, 032001 (2010).

\bibitem{Aaij:2011jp} 
R.~Aaij {\it et al.}  [LHCb Collaboration],
Phys.\ Rev.\ D {\bf 85}, 032008 (2012).  
 
\bibitem{Gronau2010}
M.~Gronau and J.~L.~Rosner,
Phys.\ Rev.\ D {\bf 83}, 034025 (2011).
  
\bibitem{bigi2011}
I.~I.~Bigi, T.~Mannel and N.~Uraltsev,
JHEP {\bf 1109}, 012 (2011).

\bibitem{charmSL}
D.~M.~Asner {\it et al.}  [CLEO Collaboration],
Phys.\ Rev.\ D {\bf 81}, 052007 (2010).

\bibitem{babarbsxlnu2012}
J.~P.~Lees {\it et al.}  [BaBar Collaboration],
Phys.\ Rev.\ D {\bf 85}, 011101 (2012).

\bibitem{Abazov:2007wg}
V.~M.~Abazov {\it et al.}  [D0 Collaboration],
Phys.\ Rev.\ Lett.\  {\bf 102}, 051801 (2009).  

\bibitem{Aaij:2011ju}
R.~Aaij {\it et al.}  [LHCb Collaboration],
Phys.\ Lett.\ B {\bf 698}, 14 (2011).  

\bibitem{KEKB}
S.~Kurokawa and E.~Kikutani, 
Nucl.\ Instrum.\ Meth.\ A {\bf 499}, 1 (2003),
and other papers included in this volume.

\bibitem{Belle}
A.~Abashian {\it et al.},
Nucl.\ Instrum.\ Meth.\ A {\bf 479}, 117 (2002).

\bibitem{nbs}
This number was obtained by the Belle Collaboration 
with the method described in
A. Drutskoy et al. [Belle Collaboration], 
Phys.\ Rev.\ Lett. {\bf 98}, 052001 (2007).

\bibitem{evtgen}
D.~J.~Lange, 
Nucl.\ Instrum.\ Meth.\ A {\bf 462}, 152 (2001).

\bibitem{geant3}
R.~Brun {\it et al.},
GEANT 3.21 CERN Report DD/EE/84-1, (1984).

\bibitem{Bailey:2012rr} 
J.~A.~Bailey {\it et al.},  
Phys.\ Rev.\ D {\bf 85}, 114502 (2012).
Erratum-ibid.\ D {\bf 86}, 039904 (2012).

\bibitem{Chen:2011ut} 
X.~J.~Chen, H.~F.~Fu, C.~S.~Kim and G.~L.~Wang,
J.\ Phys.\ G {\bf 39}, 045002 (2012).

\bibitem{Li:2010bb} 
G.~Li, F.-L.~Shao and W.~Wang,  
Phys.\ Rev.\ D {\bf 82}, 094031 (2010).

\bibitem{PDBook}
J.~Beringer {\it et al.}  [Particle Data Group],
Phys.\ Rev.\ D {\bf 86}, 010001 (2012).

\bibitem{ISGW2}
D.~Scora and N.~Isgur,
Phys.\ Rev.\ D {\bf 52}, 2783 (1995).
  
\bibitem{HQET2}
I.~Caprini, L.~Lellouch and M.~Neubert,
Nucl.\ Phys.\ B {\bf 530}, 153 (1998).

\bibitem{hfag} 
Y.~Amhis {\it et al.}  [Heavy Flavor Averaging Group],
arXiv:1207.1158 [hep-ex].
  
\bibitem{photos}
E.~Barberio and Z.~Was,
Comput.\ Phys.\ Commun.\  {\bf 79}, 291 (1994).
  
\bibitem{CC}
Throughout this paper, the inclusion of the charge conjugate mode decay is
implied unless otherwise stated.

\bibitem{Bs2DsX}
We interpret the average $\mathcal{B}(B_s^0 \to D_s^\pm X) = (93 \pm 25)\%$ 
from Ref. [16] as a multiplicity (\textit{i.e.}, the value can be greater than one), 
not as a branching fraction $\mathcal{B}(B_s^0 \to D_s^- X)$. 

\bibitem{speedoflight}
Throughout this paper, the convention $c=1$ is used.
 
\bibitem{phdlouvot}
R.~Louvot, ``Study of $B_s^0$-meson production and measurement of $B_s^0$ decays
into a $D_s^{(*)}$ and a light meson in $e^+e^-$ collisions at $\sqrt{s} =
\unit[10.87]{GeV}$,'' PhD thesis \#5213, EPFL (2012).

\bibitem{Sia:2006cq} 
R.~Sia and S.~Stone, 
Phys.\ Rev.\ D {\bf 74}, 031501 (2006), 
[Erratum-ibid.\ D {\bf 80}, 039901 (2009)].

\bibitem{Drutskoy:2010an}
A.~Drutskoy {\it et al.}  [Belle Collaboration], 
Phys.\ Rev.\ D {\bf 81}, 112003 (2010).

\bibitem{BaBarB2Ds}
B.~Aubert {\it et al.}  [BABAR Collaboration],  
Phys.\ Rev.\ D {\bf 75}, 072002 (2007).  

\end{thebibliography}
\end{document}